# Can Machine Learning discover the determining factors in participation in insurance schemes?

## A comparative analysis

Biagini Luigi[1] ,Severini Simone[1],

[1]Depart. of Agriculture and Forest Sciences (DAFNE), University of Tuscia, Viterbo, Italy

## Abstract

Identifying factors that affect participation is key to a successful insurance scheme. This study's challenges involve using many factors that could affect insurance participation to make a better forecast.Huge numbers of factors affect participation, making evaluation difficult. These interrelated factors can mask the influence on adhesion predictions, making them misleading.This study evaluated how 66 common characteristics affect insurance participation choices. We relied on individual farm data from FADN from 2016 to 2019 with type 1 (Fieldcrops) farming with 10,926 observations.We use three Machine Learning (ML) approaches (LASSO, Boosting, Random Forest) compare them to the GLM model used in insurance modelling. ML methodologies can use a large set of information efficiently by performing the variable selection. A highly accurate parsimonious model helps us understand the factors affecting insurance participation and design better products.ML predicts fairly well despite the complexity of insurance participation problem. Our results suggest Boosting performs better than the other two ML tools using a smaller set of regressors. The proposed ML tools identify which variables explain participation choice. This information includes the number of cases in which single variables are selected and their relative importance in affecting participation.Focusing on the subset of information that best explains insurance participation could reduce the cost of designing insurance schemes.

## 1  Background

External shocks, such as extreme events in weather conditions, markets or policy, significantly impact agriculture. Farmers use various risk management tools to deal with these risks, where insurance takes the lion's share (Finger et al., 2022).

The agricultural insurance literature (f.e., Meuwissen, Mey and van Asseldonk (2018) has analysed several aspects that affect the relationship between farmers and insurance. In particular, El Benni, Finger and Meuwissen (2016) emphasised the character of the variables selection and accuracy of prediction. Furthermore, the complementary effects of farm-specific characteristics and risk management strategies regarding both farm income and household income risk are analysed in El Benni, Finger and Mann (2012a) and Trestini et al. (2018), while the role of subsidies and the farm size in the stabilisation of farm income is evidenced in Aleksandrova, Zhmykhova and Viira, (2022). Moreover, Zubor-Nemes et al. (2018) highlighted the correlation between the economic performances of crop-producing farms with agricultural insurance contracts, finding that these farms outperform those who do not employ this risk-management instrument.

## 2  Topic and objectives of the analysis

We explore the elements that affect farmers' participation in an insurance scheme. This assessment is required to build or modify the structure of the insurance contract or to meet special insurance requirements based on the unique features of individual farms. Hence, this can support insurance companies and policymakers in creating contracts that satisfy farmers' needs.

This study analyses the (many) characteristics that potentially affect farmers' behaviour when considering participating in an insurance scheme using different Machin Learning tools. The number of characteristics influencing participation choice is usually large, making the task challenging. The additional problem is that these factors are interrelated and can mask the influence in the prediction of adhesions by misleading the forecast. Performing an accurate prediction and recognising the factors that affect farmers' participation are the main objectives of this analysis. Unfortunately, traditional methodologies (GLM) cannot satisfactorily use this large set of variables because of problems such as multicollinearity and overfitting. Problems that ML tools could overcome.

## 3  Methodology and data

The analysis uses individual data from the Italian FADN from 2016 to 2019. To have a homogeneous group of farms, we focus on field crop farms (type of farming 1), yielding 10,926 observations.Focusing only on one type of farm, we analyse the homogeneous class of farm insurances that covers crop production risk. To evaluate the participation in insurance schemes, we have focused on insurance subsidised by the Rural Development Program (RDP). Then, we identify the dichotomous dependent variable, taking the value of 1 when the farm buys subsidised insurance and zero otherwise.

We use 66 characteristics that the literature commonly considers to affect insurance participation choices. In particular, we consider economic, technical, financial, topographic and climatic characteristics (see f.e., (Mishra and El-Osta, 2001; Yee, Ahearn and Huffman, 2004; El Benni, Finger and Mann, 2012b; El Benni, Finger and Meuwissen, 2016; Severini, Tantari and Di Tommaso, 2016).

Because the number of participants is small (i.e., around 4% of the observations), we recur to simultaneous over- and under-sampling to create a valuable dataset for the estimation (Menardi and Torelli, 2014).

We use three Machine Learning (ML) approaches that are: LASSO, Boosting and Random Forest (Hastie, Tibshirani and Friedman, 2009; Storm, Baylis and Heckelei, 2020) to explore the issue and compare the results from these with those derived from a GLM model that has been traditionally used in insurance assessment. The considered ML approaches use a large set of variables by selecting the variable.

The ML approaches are analysed considering the following aspects: goodness-of-fit, ability to perform variable selection, and performing in variables setting. To compare the goodness-of-fit, we use Confusion Matrix analysis and metrics to compare predicted and observed values (MAE (Mean Absolute Error), MSE (Mean Squared Error), and RMSE (Root Mean Squared Error)). Moreover, one analyses the performance in variables selections, collinearity treatment and the ease-of-use (requirement of tuning).

Variable selection can be explored in two ways. First, consider the cases in which the single variables are selected. Second, consider the relative importance of each variable in affecting participation. All these results could be helpful in practice because focusing only on the subset of information that is more valuable to explain insurance participation could reduce the cost of gathering and processing information and related costs.

## 4   Results

The main preliminary results are summarised qualitatively in the following table that allows to compare the ML approaches and the GLM models.

| | | | GLM | | LASSO | | Boosting | | Random Forest |
|---|---|---|---|---|---|---|---|---|---|
| **Goodness-of-fit** | Confusion Matrix Analysis | AUC | ⬇ | 0.694 | ⬇ | 0.710 | ⬆ | 0.886 | ⬆ | 0.945 |
| | | Accuracy | ⬇ | 0.854 | ⬇ | 0.864 | ⬆ | 0.952 | 🟡 | 0.894 |
| | | Sensitivity | ⬇ | 0.428 | ⬇ | 0.454 | ⬆ | 0.776 | ⬇ | 0.491 |
| | | Specificity | ⬇ | 0.959 | ⬇ | 0.965 | ⬆ | 0.995 | ⬆ | 0.994 |
| | | Posit. Prediction Value | ⬇ | 0.722 | ⬇ | 0.763 | ⬆ | 0.977 | ⬆ | 0.950 |
| | | Negat. Prediction Value | ⬇ | 0.872 | ⬇ | 0.878 | ⬆ | 0.947 | ⬆ | 0.888 |
| | | Detection Rate | ⬇ | 0.117 | ⬇ | 0.118 | ⬆ | 0.157 | ⬇ | 0.102 |
| | | Balanced Accuracy | ⬇ | 0.694 | ⬇ | 0.710 | ⬆ | 0.886 | ⬇ | 0.742 |
| | Metrics | MAE | ⬇ | 0.146 | ⬇ | 0.136 | ⬆ | 0.048 | ⬇ | 0.149 |
| | | MSE | ⬇ | 0.146 | ⬇ | 0.136 | ⬆ | 0.048 | ⬆ | 0.079 |
| | | RMSE | ⬇ | 0.382 | ⬇ | 0.369 | ⬆ | 0.219 | 🟡 | 0.281 |
| **Selection of Variables** | | | ⬇ | 66 | ⬇ | 64 | ⬆ | 41 | ⬇ | 66 |
| **Treatment of Collinearity** | | | ⬇ | | ⬆ | | 🟡 | | 🟡 | |
| **Automatic (requires little tuning)** | | | ⬆ | | ⬆ | | 🟡 | | 🟡 | |

| **Legend** | ⬆ | Good | 🟡 | Fair | ⬇ | Poor |
|---|---|---|---|---|---|---|

Regarding the goodness-of-fit, one found Boosting overcame the performance of Random Forest and that, in turn, outperforms Lasso and GLM, which perform very poorly in predicting the number of farmers joining the subsidised insurance scheme. Despite the over-under sample techniques, the number of positive values is not detected in the same way by Boosting and Random Forest, with the latter resulting in poor performance (Sensitivity, Negative Prediction Value, Detection Rate and Balanced Accuracy). MAE, MSE and RMSE confirm the best Boosting performance, the poor outcomes reached by GLM and LASSO, and finally, Random Forest shows mixed results. Boosting also prevails in selecting variables: this model can reach high performance using only 41 variables on 66. Other models present low capacity in selection variables. These models present different capacities to fight collinearity, with LASSO as the best performers, followed by Boosting and Random Forest. Moreover, the powerlessness of GLM to select variables makes this model off the comparison. Finally, we must draw attention to various difficulties encountered while setting up these instruments: Contrary to Boosting and Random Forest, where it is essential to pay attention to specific non-automatic processes, GLM and LASSO do not provide the need for tuning.

Two additional aspects are under investigation, and extensive results will be provided in the full version of the paper: i) Which variables are selected the most? ii) Which variables are the most important? Preliminary results show that the most important factors that affect insurance participation (in order of importance) are: farm economic size, presence of other gainful activities, amount of utilised agricultural area, kW of available machinery, production diversification (Herfindahl index), degree of intensification (as total revenue per unit of utilised agricultural area), fixed capital on total capital, and mechanical expenses.

## 5    Discussion

Although participation in an insurance scheme is a complex decision, ML ensures relatively good prediction for sure better than GLM models. Within the considered ML approaches, Boosting offers better performances in this regard than the other two considered ML tools. Furthermore, it also uses a smaller set of variables as regressors. Conversely, the setting of Boosting can be challenging, and the evaluation of trade-offs with performance must be necessary to consider the different variables utilised in the estimation. The proposed ML tools allow identifying the essential variables in explaining participation choice. The general conclusion is that ML is a helpful tool for exploring the factors that explain farmers' participation in insurance schemes. Furthermore, results obtained using these approaches can be useful to better design insurance schemes and, hopefully, boost farmers' participation. Therefore, the ML approach is a key step that should be done carefully considering the characteristics of the empirical case study.